\documentclass{moriond}
\usepackage{graphicx} 

\usepackage{mathtools}
\usepackage{amsmath}
\usepackage{amssymb}

\newcommand{\Photo}{}

\begin{document}
\vspace*{4cm}
\title{Geodesic Completeness in General Cosmological Scenarios}

\author{William H. Kinney}

\address{Department of Physics, University at Buffalo, SUNY\\
    Buffalo, NY 14260 USA}

\maketitle

\abstract{The well-known Borde-Guth-Vilenkin Theorem shows that inflationary spacetimes are generically geodesically past-incomplete, necessitating the existence of a pre-inflationary boundary of some sort, possibly singular. I discuss the generalization of the BGV theorem to spacetimes beyond inflation, including inhomogeneous and cyclic models. As an example, I show that the cyclic model proposed by Ijjas and Steinhardt is geodesically incomplete. 
}

\section{Geodesic Completeness in FRW spacetime} \label{sec:BGVTheorem}

Under what circumstances is a spacetime finite to the past? It is clear that power-law evolution of the scale factor, 
\begin{equation}
    a\left(t\right) \propto t^{\alpha},\quad \alpha > 0,
\end{equation}
results in a universe with finite age, since $a \to 0$ as $t \to 0$. Such evolution follows from a constant equation of state $p = w \rho$ with $w = \text{const.},\ w > -1$. The existence of an initial singularity is not so clear in the case of de Sitter evolution, with
\begin{equation}
    a\left(t\right) \propto e^{H t},\quad H = \text{const.}
\end{equation}
since $a \to 0$ only when $t \to -\infty$. At first glance, this suggests that inflation originating from a de Sitter phase might be extended infinitely into the past, neatly solving the problem of an initial singularity in cosmology. However, the coordinate time $t$ corresponds to the time as measured by a comoving observer, and is not invariant under coordinate transformations. We need a way to categorize spacetime as bounded or unbounded in a way that does not depend on the choice of coordinate system. This is the idea of \textit{geodesic completeness}: a spacetime is geodesically complete if all possible geodesic world lines extend infinitely into both the past and the future as measured by the proper time along the world line, 
\begin{equation}
    \int_{-\infty}^t{d s} \to \infty, \qquad \int_t^{+\infty}{ds} \to \infty. 
\end{equation}
For exponential expansion, comoving geodesics extend infinitely into both the past and the future, but that is not sufficient to demonstrate geodesic completeness, which requires this to be true for \textit{all} geodesics.

We consider the case of a general Friedmann-Robertson-Walker (FRW) spacetime. A theorem introduced by Borde, Guth, and Vilenkin (the \textit{BGV theorem}) \cite{Borde:2001nh} states that any spacetime for which the average expansion rate $H_{\text{av}}$ is positive is geodesically incomplete, where
\begin{equation} \label{eq:Hav}
    H_{\text{av}} \equiv \frac{1}{\Delta s} \int{H ds},
\end{equation}
and $ds$ is the proper time along a geodesic world line. First, let us consider the integral of the expansion rate $H$ along a comoving world line, for which the proper time $ds = dt$,
\begin{equation} \label{eq:Hdt}
    \begin{aligned}
        \int_{t_i}^{t_f}{H dt} &= \int_{t_i}^{t_f}{\frac{1}{a} \frac{d a}{d t} dt} = \int_{a_i}^{a_f}{d \ln{a}} = \ln{\left(a_f\right)} - \ln{\left(a_i\right)}. 
    \end{aligned}
\end{equation}
Since the integrand is a total derivative, the integral is determined by the values of the scale factor on the boundary, and measures the total expansion from $t_i$ to $t_f$. This integral diverges logarithmically as $a_i \to 0$; in the de Sitter case, this corresponds to infinite past coordinate time. 

For a general timelike geodesic with four-velocity $u^\mu$ tangent to the geodesic world line, normalization $u^\mu u_\mu = -1$ gives
\begin{equation}
    u^\mu u_\mu = g_{\mu \nu} \frac{d x^\mu}{ds} \frac{d x^\nu}{ds} = - \left(\frac{d t}{ds}\right)^2 + a^{2}\left(t\right) \left\vert \frac{d {\vec x}}{d s}\right\vert^2 = -1.
\end{equation}
The geodesic equation governing a freely falling observer is
\begin{equation}
    \frac{d}{d s}\left[a^2\left(t\right) \frac{d {\vec x}}{d s}\right] = 0,
\end{equation}
so that we can define an integration constant $v_0$ such that \cite{Aguirre:2001ks}
\begin{equation}
a^2\left(t\right) \left\vert \frac{d {\vec x}}{d s} \right\vert \equiv v_0 = \text{const.}
\end{equation}
We then have
\begin{equation} \label{eq:dtdstimelike}
    \left(\frac{d t}{ds}\right)^2 = \gamma^2 = 1 + v_0^2 a^{-2}\left(t\right),
\end{equation}
where
\begin{equation}
    \gamma^2 = \frac{1}{1 - v^2},
\end{equation}
is the Lorentz boost as a function of the three-velocity $v$ of the observer. We can integrate the expansion rate $H$ along a non-comoving geodesic with proper time $d s$ as follows:
\begin{equation}
    \begin{aligned}
        \int_{s_i}^{s_f} H ds &= \int_{t_i}^{t_f} H \frac{d s}{d t} dt \\
                              &= \int_{t_i}^{t_f}{\frac{1}{a}\left(\frac{d a}{d t}\right) \frac{1}{\sqrt{1 + v_0^2 / a^2}} dt} = \int_{a_i}^{a_f}{\frac{d a}{\sqrt{a^2 + v_0^2}}} \\
                              &= \frac{1}{2} \left.\ln{\left[\frac{\sqrt{v_0^2 + a^2} + a}{\sqrt{v_0^2 + a^2} - a}\right]} \right\vert_{a_i}^{a_f}. 
    \end{aligned} 
\end{equation}
Integrating from $a_i = 0$ to $a_f = 1$, this integral is finite:
\begin{equation} \label{eq:BGVRelation}
    \begin{aligned}
        \int_{s_i}^{s_f} H ds &= \frac{1}{2} \ln \left(\frac{\sqrt{1 + v_0^2} + 1}{\sqrt{1 + v_0^2} - 1}\right) = \frac{1}{2} \ln \left(\frac{\gamma_0 + 1}{\gamma_0 - 1}\right).
    \end{aligned}
\end{equation}
We can then define an average expansion rate
\begin{equation}
    H_{\text{av}} \equiv \frac{1}{\Delta s} \int{H ds}. 
\end{equation}
Since the average expansion $H_{\text{av}}$ along the timelike world line is finite and positive, the corresponding proper time $\Delta s$ must also be finite,
\begin{equation} \label{eq:BGV}
    \Delta s = \int{d s} = \frac{1}{H_{\text{av}}} \int{H ds}, 
\end{equation}
and the spacetime is geodesically incomplete. 

Care is needed when formulating a precise statement of the theorem. This can be seen from considering the example of a ``loitering'' universe, 
\begin{equation} \label{eq:Loitering}
    a\left(t\right) = a_i \left(1 + e^{H t}\right). 
\end{equation}
Unlike the de Sitter case, the scale factor asymptotically approaches a finite value $a \to a_i$ as $t \to -\infty$; because it approaches Minkowski space at early times, it is geodesically past-complete, despite being monotonically expanding \cite{Ellis:2002we,Lesnefsky:2022fen}. This appears to be a counterexample to the BGV theorem, since the Hubble parameter is everywhere finite and positive, yet the space is geodesically complete. Such loitering models can evade the BGV bound precisely because their early-time behavior asymptotically mimics Minkowski space. To see why it is not a counterexample, consider the integral
\begin{equation} 
    \int_{-\infty}^{t_0}{H dt} = \ln{\left(a_f\right)} - \ln{\left(a_i\right)} = \ln{\left(1 + e^{H t_0}\right)}. 
\end{equation}
This is finite and positive. However, since the interval in question is infinite as $\Delta t \to \infty$, the average Hubble parameter vanishes for any $t_0$:
\begin{equation}
   H_{\text{av}} = \lim_{t_i \to -\infty} \frac{1}{\Delta t} \int_{t_i}^{t_0}{H dt} \to 0. 
\end{equation}
Contrast this with the case of de Sitter evolution with $H = \text{const.}$,
\begin{equation}
    H_{\text{av}} = \lim_{t_i \to -\infty} \frac{1}{\Delta t} \int_{t_i}^{t_0}{H dt} =  H > 0, 
\end{equation}
so that, even though the integral (\ref{eq:Hdt}) is divergent, the average (\ref{eq:Hav}) along the interval is finite and positive, and the spacetime is geodesically incomplete. 

A physical interpretation of the BGV theorem is that $H_{\text{av}}$ over an interval depends only on the \textit{amount} of expansion on that interval, not the specific form of $H\left(t\right)$, as can be seen from Eq.~(\ref{eq:Hdt}). For a given expansion history $H\left(t\right)$, assume we can define a bounding de Sitter space with $H  = H_{\text{av}} = \text{const.}$ such that 
\begin{equation}
    H_{\text{av}} \int_{t_i}^{t_f} dt = \int_{t_i}^{t_f} H dt, 
\end{equation}
and the ratio $a_f / a_i$ on the interval $\left(t_i, t_f\right)$ is therefore the same for both spaces. We then choose a scale factor $a_0$ in the bounding de Sitter space such that
\begin{equation}
    a_0 e^{H_{\text{av}} t} \geq a(t),\quad \forall t \in \left(t_i,t_f\right).
\end{equation}
The past proper time along a geodesic in the target space is then bounded from above by the past proper time in the bounding de Sitter space,
\begin{equation}
    \Delta s = \int_{t_i}^{t_f}{\frac{d t}{\sqrt{1 + v_0^2 a^{-2}\left(t\right)}}} \leq \int_{t_i}^{t_f}{\frac{d t}{\sqrt{1 + v_0^2 a_0^{-2} e^{-2 H_{\text{av}} t}}}}.
\end{equation}
Thus, geodesic incompleteness of the bounding de Sitter space implies geodesic incompleteness of the target space with expansion history $H\left(t\right)$. No such bounding space exists for the loitering cosmology (\ref{eq:Loitering}), so the theorem does not apply. In a flat background, the scale factor (\ref{eq:Loitering}) also has ${\dot H} > 0$, and therefore violates the Null Energy Condition, since ${\dot H} > 0$ requires $p < -\rho$. This suggests that a rigorous statement of geodesic incompleteness will implicitly include the assumption that the Null Energy Condition is satisfied, $p \geq -\rho$. We next consider an example of the construction of such a bounding space.  

\section{The Ijjas-Steinhardt Cyclic Model is Geodesically Incomplete}

It has been known since work by Tolman in the 1930s that cyclic cosmology suffers from the problem of entropy growth \cite{Tolman:1931fei}. The second law of thermodynamics operates identically whether the universe is expanding or contracting, and the growth of entropy with time prevents cyclic behavior. Black hole formation is a particularly difficult problem: black holes formed during one period of expansion and contraction inevitably lead to a strongly inhomogeneous bounce, regardless of the background dynamics. A model proposed by Ijjas and Steinhardt (IS) \cite{Ijjas:2019pyf,Ijjas:2021zwv} suggests a solution to the entropy problem in cyclic cosmology. In this model, the Hubble parameter for the expansion rate is periodic in time, but the cosmic scale factor grows exponentially from one oscillation to the next, diluting entropy between cycles. 

We consider a cosmology with alternating expanding and contracting phases with period $T$, such that the Hubble parameter is exactly periodic,
\begin{equation}
    H\left(t + T\right) = H\left(t\right). 
\end{equation}
The scale factor, however, grows exponentially from one cycle to the next, as a mechanism for shedding the entropy produced in each cycle,
\begin{equation}
    a\left(t + T\right) = e^N a\left(t\right),
\end{equation}
where $N$ is a constant chosen to be large enough to ensure a sufficiently smooth boundary condition at the bounce.  Calling $H_+\left(t\right) > 0$ the Hubble parameter during expansion, and $H_-\left(t\right) < 0$ the corresponding rate of contraction, the condition for entropy dissipation is 
\begin{equation}
    \int_0^\tau{H_+\left(t\right) dt} \gg - \int_\tau^T{H_-\left(t\right) d t},
    \label{eq:EntropyCondition}
\end{equation}
where $\tau$ is the duration of the expanding phase. Despite periodic $H\left(t\right)$, a difference in average equation of state between expanding and contracting phases results in growth of the scale factor. A simplified example is an expanding phase dominated by a cosmological constant $w_+ = -1$, so that $H_+ = \mathrm{const.}$, and $a_+\left(t\right) = e^{H t}$. The contracting phase is dominated by a scalar field $\phi$ with negative potential, so that
\begin{equation}
    \rho_- = \frac{1}{2} \dot\phi^2 + V\left(\phi\right) \simeq 0.
\end{equation}
Approximately vanishing density means that $H_- \simeq 0$, with the result that $a_-\left(t\right) \simeq \mathrm{const.}$ This is the well-studied case of ``ekpyrotic'' contraction (for a review, see  Ref. \cite{Lehners:2008vx}.) In this limit,
\begin{equation}
    N_+ = \int_0^\tau{H_+\left(t\right) dt} \simeq H_+ \tau,
\end{equation}
and
\begin{equation}
    N_- = \int_\tau^T{H_-\left(t\right) dt} \simeq 0,
\end{equation}
which trivially satisfies the general condition (\ref{eq:EntropyCondition}). In this limit, entropy is diluted during a period of late exponential expansion (inflation), with a subsequent slow contraction phase during which the scale factor remains approximately constant. 

\begin{figure}
    \centering
    \includegraphics[width=0.6\columnwidth]{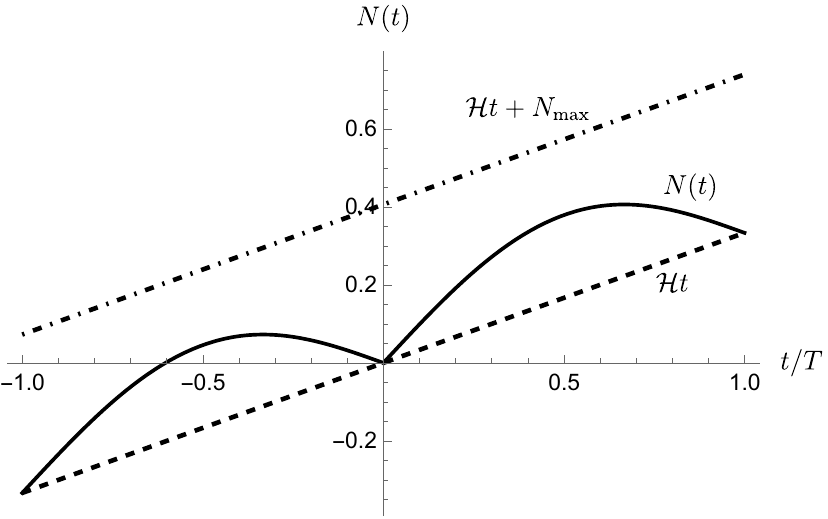}
    \caption{The number of e-folds $N\left(t\right) = \int{H\left(t\right) d t}$ vs time for the cyclic case (solid line), and the bounding de Sitter solutions (dashed, dotted).}
    \label{fig:Nvst}
\end{figure}

We consider a universe with $H(t)$ periodic with period $T$, which satisfies the condition (\ref{eq:EntropyCondition}) for entropy dissipation, $N_+ + N_- > 0$. We assume that the bounce is rapid enough that we can take $H\left(t\right)$ to be discontinuous across the bounce at the boundary of the interval, with scale factor $a\left(t\right)$ continuous.  The condition for entropy dissipation (\ref{eq:EntropyCondition}) means there must exist a mean expansion rate $\mathcal{H}$ on the interval $0 < t < T$ such that
\begin{equation}
    \mathcal{H} \equiv \frac{1}{T} \int_{0}^T{H\left(t'\right) dt'} > 0. 
\end{equation}
Requiring that the Null Energy Condition holds everywhere except the bounce is equivalent to the condition that $H\left(t\right)$ be constant or decreasing,
\begin{equation}
    \frac{d H\left(t\right)}{d t} \leq 0,
    \label{eq:NEC}
\end{equation}
for $t$ in the interval $0 < t < T$. Then
\begin{equation}
    e^{\mathcal{H} t} < \exp\left[\int_{0}^t{H\left(t'\right) dt'}\right]
\end{equation}
for all $0 < t < T$. The Null Energy Condition (\ref{eq:NEC}) likewise requires that there exist a finite $N_{\mathrm{max}}$
\begin{equation}
    N_{\mathrm{max}} \equiv \mathrm{Max}\left[\int_{0}^t{H\left(t'\right) dt'}\right],
\end{equation}
such that
\begin{equation}
    \int_{0}^t{H\left(t'\right) dt'} < \mathcal{H} t + N_{\mathrm{max}}
\end{equation}
for all $0 < t < T$. The number of e-folds of expansion is then bounded from above and below by two de Sitter solutions (Fig. \ref{fig:Nvst}),
\begin{equation}
    e^{\mathcal{H} t} < \exp\left[\int_{0}^t{H\left(t'\right) dt'}\right] < e^{\mathcal{H} t + N_{\mathrm{max}}}.
    \label{eq:Nefoldsbound}
\end{equation}
The proper time of a geodesic path on any \textit{past} interval integrating from $t = -t_i$ to $t = 0$ is given by
\begin{equation}
    \Delta s = \int_{-t_i}^{0}{\frac{dt}{\sqrt{1 + v_0^2 a^{-2}\left(t\right)}}},
\end{equation}
with 
\begin{equation}
    a\left(t\right) =  \exp\left[{\int_0^t{H\left(t'\right) dt'}}\right]. 
\end{equation}
From Eq. (\ref{eq:Nefoldsbound}), proper time satisfies a bound $\Delta s_{\mathrm{min}} < \Delta s < \Delta s_{\mathrm{max}}$, where
\begin{equation}
    \Delta s_{\mathrm{min}} = \int_{-t_i}^{0}{\frac{dt}{\sqrt{1 + v_0^2 e^{-2 \mathcal{H} t}}}} = \frac{1}{2 \mathcal{H}} \ln \left(\frac{\sqrt{1 + v_0^2} + 1}{\sqrt{1 + v_0^2} - 1}\right).
\end{equation}
and
\begin{equation}
    \Delta s_{\mathrm{max}} = \int_{-t_i}^{0}{\frac{dt}{\sqrt{1 + v_0^2 e^{-2 N_{\mathrm{max}}}  e^{-2 \mathcal{H} t}}}}= \frac{1}{2 \mathcal{H}} \ln \left(\frac{\sqrt{1 + v_0^2 e^{-2 N_{\mathrm{max}}}} + 1}{\sqrt{1 + v_0^2e^{-2 N_{\mathrm{max}}} } - 1}\right).
\end{equation}
We then have that $\Delta s_{\mathrm{min}}$ and $\Delta s_{\mathrm{max}}$ are finite on the interval $-\infty < t < 0$, which means that $\Delta s_{\mathrm{min}} < \Delta s < \Delta s_{\mathrm{max}}$ must also be finite on the same interval. Any spacetime periodic in $H\left(t\right)$ and satisfying the condition $(\ref{eq:EntropyCondition})$ must therefore be geodesically incomplete.

\section{BGV Theorem: General Case} \label{sec:BGVGeneral}

To extend the applicability of the BGV theorem, we now consider arbitrary spacetimes without imposing FRW symmetry. Take a timelike geodesic congruence $\left\lbrace u^\mu \right\rbrace$. The condition for $u^\mu$ to be a geodesic is
\begin{equation}
    \frac{d u^\mu}{d s} = u^\nu u^\mu{}_{;\nu} = 0,
\end{equation}
where as usual a semicolon indicates a covariant derivative. The expansion rate relative to $\left\lbrace u^\mu \right\rbrace$ is defined by the divergence of the four-velocity
\begin{equation}
    \Theta \equiv u^\mu{}_{;\mu}. 
\end{equation}
We can write the four-divergence $\Theta$  as
\begin{equation}
    \Theta = \lambda^{\mu \nu} \nabla_\mu u_\nu = \lambda^{\mu \nu} K_{\mu \nu},
\end{equation}
where
\begin{equation}
    \lambda^{\mu \nu} \equiv g^{\mu \nu} + u^\mu u^\nu
\end{equation}
 is the spatial projection tensor, and $K_{\mu \nu}$ is the extrinsic curvature of hypersurfaces orthogonal to $u^\mu$. We can then define a generalized version of $H_{\text{av}}$ (\ref{eq:Hav}) in terms of the extrinsic curvature as
\begin{equation} \label{eq:BGVBasicIntegral}
    \Theta_{\text{av}} \equiv \frac{1}{\Delta s} \int{\Theta d s} = \frac{1}{\Delta s} \int{\lambda^{\mu \nu} K_{\mu \nu} \ ds},
\end{equation}
where $ds$ is the proper time along $u$. We can evaluate this integral by writing the argument in terms of the Lie derivative of the induced metric $h_{i j}$ on spatial hypersurfaces,
\begin{equation}
    \lambda^{\mu \nu} K_{\mu \nu} = - \frac{1}{2} h^{i j} \left[\mathcal{L}_u h_{i j}\right] = \frac{1}{2} \frac{d}{d s} Tr\left[\ln{\left(h_{i j}\right)}\right],
\end{equation}
where $\mathcal{L}_u$ is the Lie derivative along $u^\mu$. 

As an example, we can take a flat FRW metric,
\begin{equation}
    g_{\mu \nu} = \text{diag.}\left[-1, a^2(t), a^2(t), a^2(t)\right].
\end{equation}
Taking $u^\mu$ to be comoving world lines, the induced metric is then
\begin{equation}
    h_{i j} = \text{diag.}\left[a^2(t), a^2(t), a^2(t)\right],
\end{equation}
and
\begin{equation}
    - \frac{1}{2} h^{i j} \left[\mathcal{L}_u \left(h_{i j}\right)\right] =  \frac{3}{2 a^2} \frac{d}{d t} \left(a^2\right) = 3 \frac{\dot a}{a} = 3 H.
\end{equation}
This is just the usual Hubble parameter, and we have
\begin{equation}
    \int{\Theta ds} = 3 \int{H dt} = \frac{1}{2} \text{Tr}\left[\ln{(h_{i j})}\right] = 3 \ln{a},
\end{equation}
which is just Eq.~(\ref{eq:Hdt}), where the log of the trace of the induced metric is just the log of the scale factor.

We can now construct a general version of the BGV Theorem. Take a geodesic congruence $\left\lbrace u^\mu \right\rbrace$ with zero shear and vorticity $\sigma_{\mu \nu} = \omega_{\mu \nu} = 0$. Then choose a geodesic $\mathcal{C}$ with four-velocity $\left\lbrace v^\mu\right\rbrace \neq \left\lbrace u^\mu\right\rbrace$. We can define
\begin{equation}
    \gamma \equiv - v^\mu u_\mu,
\end{equation}
where $\gamma$ can be identified as the Lorentz boost of the vector $v^\mu$ relative to the rest frame $u^\mu = (1, 0, 0, 0)$. If $v^\mu$ is timelike, its normalization is
\begin{equation} \label{eq:normalization}
    \begin{aligned}
        v^\mu v_\mu &= -1 = v^\mu \left(g_{\mu \nu} v^\nu\right) \\
                    &= v^\mu \left(\lambda_{\mu \nu} - u_\mu u_\nu\right) v^\nu \\
                    &= - \gamma^2 + \lambda_{\mu \nu} v^\mu v^\nu. 
    \end{aligned}
\end{equation}
Normalization of $v^\mu$ therefore results in the relation
\begin{equation}
    \lambda_{\mu \nu} v^\mu v^\nu = \gamma^2 - 1. 
\end{equation}
We also have the geodesic equation,
\begin{equation}
    \frac{d v^\mu}{d s} = v^\nu v_{\mu;\nu} = 0,
\end{equation}
where $ds$ is the proper time measured along $v^\mu$. Then
\begin{equation}
    \begin{aligned}
        \frac{d \gamma}{d s} &= - \frac{d}{d s} \left(v^\mu u_\mu\right)\\
                             &= - v^\mu v^\nu u_{\mu;\nu} = - \frac{1}{3} \Theta v^\mu v^\nu \lambda_{\mu \nu} \\
                             &= \frac{1}{3} \Theta \left(1 - \gamma^2\right),
    \end{aligned}
\end{equation}
We then have the relation
\begin{equation}
    \Theta = \frac{3}{1 - \gamma^2} \frac{d \gamma}{d s}.
\end{equation}
Integrating with respect to proper time along the geodesic $\mathcal{C}$,
\begin{equation} \label{eq:FTimelike}
    \begin{aligned}
        \int{\Theta ds} &= \int{\frac{3}{1 - \gamma^2} \frac{d \gamma}{d s} ds} \\
                        &= \int{\frac{3 d\gamma}{1 - \gamma^2}} \\
                        &= \frac{3}{2} \ln\left(\frac{\gamma + 1}{\gamma - 1}\right).
    \end{aligned}
\end{equation}
Similarly, for a null geodesic $k^\mu$, define
\begin{equation}
    \gamma \equiv k^\mu u_\mu.
\end{equation}
Here $\gamma$ is not a Lorentz boost as in the timelike case, but the contraction of the null vector with the timelike congruence $u^\mu$. Normalization implies
\begin{equation}
    k^\mu k_\mu = 0 = - \gamma^2 + \lambda_{\mu \nu} k^\mu k^\nu,
\end{equation}
and the spatial projection is
\begin{equation}
    \lambda_{\mu \nu} k^\mu k^\nu = \gamma^2. 
\end{equation}
We can write the geodesic equation for $k^\mu$ in terms of an affine parameter $\alpha$ as
\begin{equation}
    \frac{d k^\mu}{d \alpha} = k^\nu k_{\mu;\nu} = 0. 
\end{equation}
For the null case, the derivative of $\gamma$ with respect to the affine parameter satisfies
\begin{equation}
    \begin{aligned}
        \frac{d \gamma}{d \alpha} &= - \frac{d}{d \alpha} \left(k^\mu u_\mu\right)\\
                             &= - k^\mu k^\nu u_{\mu;\nu} = - \frac{1}{3} \Theta k^\mu k^\nu \lambda_{\mu \nu} \\
                             &= - \frac{1}{3} \gamma^2 \Theta.
    \end{aligned}
\end{equation}
As in the timelike case, we can then write the expansion parameter as
\begin{equation}
    \Theta = - \frac{3}{\gamma^2} \frac{d \gamma}{d \alpha},
\end{equation}
and
\begin{equation} \label{eq:FNull}
    \begin{aligned}
        \int{\Theta d\alpha} &= - \int{\frac{3}{\gamma^2} \frac{d \gamma}{d \alpha} d\alpha} \\
                        &= -\int{\frac{3 d\gamma}{\gamma^2}} = \frac{3}{\gamma}. 
    \end{aligned}
\end{equation}

A rigorous statement of the BGV theorem is as follows \cite{Pavlovic:2023mke,Garcia-Saenz:2024ogr}:

\bigskip
\textbf{Theorem:}
    Consider a geodesic $\mathcal{C}$ parameterized by affine parameter $\alpha$. If there exists a constant $\Delta > 0$ such that
    \begin{equation} 
        \Theta_{\text{av}} \equiv \frac{1}{\alpha_f - \alpha_0} \int_{\alpha_0}^{\alpha_f}{\Theta d\alpha} \geq \Delta\quad \forall \alpha_0 \in \left(\alpha_i,\alpha_f\right),
    \end{equation}
    then $\mathcal{C}$ is past-incomplete.

\bigskip
\textbf{Proof}
    We use $\alpha$ as a general affine parameter, which reduces to proper time $s$ for timelike geodesics and is affine for null geodesics. From Eqs. (\ref{eq:FTimelike}) and (\ref{eq:FNull}), 
    \begin{equation} 
        \Theta_{\text{av}} \equiv \frac{1}{\alpha_f - \alpha_0} \int_{\alpha_0}^{\alpha_f}{\Theta d\alpha} = \frac{F\left(\gamma_f\right) - F\left(\gamma_0\right)}{\alpha_f - \alpha_0} \geq \Delta,
    \end{equation}
    where $F\left(\gamma\right)$ is given by 
    \begin{equation}
        F\left(\gamma\right) = \frac{3}{2} \ln\left(\frac{\gamma + 1}{\gamma - 1}\right),
    \end{equation}
    if $\mathcal{C}$ is timelike, and 
    \begin{equation}
        F\left(\gamma\right) = \frac{3}{\gamma},
    \end{equation}
    if $\mathcal{C}$ is null. Then 
    \begin{equation}
        \Delta \leq \Theta_{\text{av}} < \frac{F\left(\gamma_f\right)}{\alpha_f - \alpha_0}\quad \forall \alpha_0 \in \left(\alpha_i,\alpha_f\right).
    \end{equation}
    Then
    \begin{equation}
        \alpha_0 > \alpha_f - \frac{F\left(\gamma_f\right)}{\Delta}\quad \forall \alpha_0 \in \left(\alpha_i,\alpha_f\right).
    \end{equation}
    Therefore $\alpha_i$ is bounded from below,
    \begin{equation}
        \alpha_i \geq \alpha_f - \frac{F\left(\gamma_f\right)}{\Delta},
    \end{equation}
    and the affine length of the geodesic $\mathcal{C}$ is finite. Therefore, the spacetime is geodesically incomplete. 

\section*{Acknowledgements}

This proceedings contribution is based on work in collaboration with Nina Stein, Suvashis Maity, and L. Sriramkumar \cite{Kinney:2021imp,Kinney:2023urn}. Material is also excerpted from my upcoming book, \textit{Inflationary Cosmology} \cite{Kinney2025InflationaryCosmology}. This work is supported by the National Science Foundation under grant NSF-PHY-2310363.

\bibliography{Moriond2026.bib}

@article{Borde:2001nh,
    author = "Borde, Arvind and Guth, Alan H. and Vilenkin, Alexander",
    title = "{Inflationary space-times are incompletein past directions}",
    eprint = "gr-qc/0110012",
    archivePrefix = "arXiv",
    reportNumber = "MIT-CTP-3183",
    doi = "10.1103/PhysRevLett.90.151301",
    journal = "Phys. Rev. Lett.",
    volume = "90",
    pages = "151301",
    year = "2003"
}

@article{Ellis:2002we,
    author = "Ellis, George F. R. and Maartens, Roy",
    title = "{The emergent universe: Inflationary cosmology with no singularity}",
    eprint = "gr-qc/0211082",
    archivePrefix = "arXiv",
    doi = "10.1088/0264-9381/21/1/015",
    journal = "Class. Quant. Grav.",
    volume = "21",
    pages = "223--232",
    year = "2004"
}

@article{Lesnefsky:2022fen,
    author = "Lesnefsky, J. E. and Easson, D. A. and Davies, P. C. W.",
    title = "{Past-completeness of inflationary spacetimes}",
    eprint = "2207.00955",
    archivePrefix = "arXiv",
    primaryClass = "gr-qc",
    doi = "10.1103/PhysRevD.107.044024",
    journal = "Phys. Rev. D",
    volume = "107",
    number = "4",
    pages = "044024",
    year = "2023"
}

@article{Pavlovic:2023mke,
    author = "Pavlovi\'c, Petar and Sossich, Marko",
    title = "{Geodesically complete cyclic cosmologies and entropy}",
    eprint = "2305.06719",
    archivePrefix = "arXiv",
    primaryClass = "gr-qc",
    doi = "10.1140/epjc/s10052-024-12621-z",
    journal = "Eur. Phys. J. C",
    volume = "84",
    number = "3",
    pages = "242",
    year = "2024"
}

@article{Garcia-Saenz:2024ogr,
    author = "Garcia-Saenz, Sebastian and Hua, Junjie and Zhao, Yunke",
    title = "{Geodesic completeness, cosmological bounces, and inflation}",
    eprint = "2405.04062",
    archivePrefix = "arXiv",
    primaryClass = "gr-qc",
    doi = "10.1103/PhysRevD.110.L061304",
    journal = "Phys. Rev. D",
    volume = "110",
    number = "6",
    pages = "L061304",
    year = "2024"
}

@article{Tolman:1931fei,
    author = "Tolman, Richard C.",
    title = "{On the Theoretical Requirements for a Periodic Behaviour of the Universe}",
    doi = "10.1103/PhysRev.38.1758",
    journal = "Phys. Rev.",
    volume = "38",
    number = "9",
    pages = "1758",
    year = "1931"
}

@article{Ijjas:2019pyf,
    author = "Ijjas, Anna and Steinhardt, Paul J.",
    title = "{A new kind of cyclic universe}",
    eprint = "1904.08022",
    archivePrefix = "arXiv",
    primaryClass = "gr-qc",
    doi = "10.1016/j.physletb.2019.06.056",
    journal = "Phys. Lett. B",
    volume = "795",
    pages = "666--672",
    year = "2019"
}

@article{Ijjas:2021zwv,
    author = "Ijjas, Anna and Steinhardt, Paul J.",
    title = "{Entropy, Black holes, and the New Cyclic Universe}",
    eprint = "2108.07101",
    archivePrefix = "arXiv",
    primaryClass = "gr-qc",
    month = "8",
    year = "2021"
}

@article{Lehners:2008vx,
    author = "Lehners, Jean-Luc",
    title = "{Ekpyrotic and Cyclic Cosmology}",
    eprint = "0806.1245",
    archivePrefix = "arXiv",
    primaryClass = "astro-ph",
    doi = "10.1016/j.physrep.2008.06.001",
    journal = "Phys. Rept.",
    volume = "465",
    pages = "223--263",
    year = "2008"
}

@article{Aguirre:2001ks,
    author = "Aguirre, Anthony and Gratton, Steven",
    title = "{Steady state eternal inflation}",
    eprint = "astro-ph/0111191",
    archivePrefix = "arXiv",
    doi = "10.1103/PhysRevD.65.083507",
    journal = "Phys. Rev. D",
    volume = "65",
    pages = "083507",
    year = "2002"
}

@article{Kinney:2021imp,
    author = "Kinney, William H. and Stein, Nina K.",
    title = "{Cyclic cosmology and geodesic completeness}",
    eprint = "2110.15380",
    archivePrefix = "arXiv",
    primaryClass = "gr-qc",
    doi = "10.1088/1475-7516/2022/06/011",
    journal = "JCAP",
    volume = "06",
    number = "06",
    pages = "011",
    year = "2022"
}

@article{Kinney:2023urn,
    author = "Kinney, William H. and Maity, Suvashis and Sriramkumar, L.",
    title = "{Borde-Guth-Vilenkin theorem in extended de Sitter spaces}",
    eprint = "2307.10958",
    archivePrefix = "arXiv",
    primaryClass = "gr-qc",
    doi = "10.1103/PhysRevD.109.043519",
    journal = "Phys. Rev. D",
    volume = "109",
    number = "4",
    pages = "043519",
    year = "2024"
}

@book{Kinney2025InflationaryCosmology,
  author    = {Will Kinney},
  title     = {Inflationary Cosmology},
  publisher = {World Scientific},
  year      = {2026},
  series    = {World Scientific Series in Cosmology and Astroparticle Physics},
  doi       = {10.1142/14698},
  url       = {https://www.worldscientific.com/worldscibooks/10.1142/14698}
}

\end{document}